# Imaging magnetism evolution of magnetite to megabar pressure range with quantum sensors in diamond anvil cell

**Authors:**

Mengqi Wang[1,3]†, Yu Wang[2,6]†*, Zhixian Liu[1,3]†, Ganyu Xu[1,3], Bo Yang[1,3], Pei Yu[1,3], Haoyu Sun[1,3], Xiangyu Ye[1,3], Jingwei Zhou[1,3], Alexander. F. Goncharov[7], Ya Wang[1,3,4]* and Jiangfeng Du[1,3,4,5]*

**Affiliations:**

[1] CAS Key Laboratory of Microscale Magnetic Resonance and School of Physical Sciences, University of Science and Technology of China, Hefei 230026, China

[2] Key Laboratory of Materials Physics, Institute of Solid State Physics, HFIPS, Chinese Academy of Sciences, Hefei, China.

[3] CAS Center for Excellence in Quantum Information and Quantum Physics, University of Science and Technology of China, Hefei 230026, China

[4] Hefei National Laboratory, University of Science and Technology of China, Hefei 230088, China

[5] Institute of Quantum Sensing and School of Physics, Zhejiang University, Hangzhou 310027, China

[6] Institute of Geosciences, Goethe University Frankfurt, Frankfurt 60438, Germany

[7] Earth and Planets Laboratory, Carnegie Institution of Washington, Washington, DC, USA.

*Corresponding author. Email: wangyu@issp.ac.cn (Yu Wang), ywustc@ustc.edu.cn (Ya Wang) and djf@ustc.edu.cn (Jiangfeng Du)

†These authors contributed equally to this work

**Abstract**

High-pressure diamond anvil cells have been widely used to create novel states of matter. Nevertheless, the lack of universal in-situ magnetic measurement techniques at megabar pressures makes it difficult to understand the underlying physics of materials' behavior at extreme conditions, such as high-temperature superconductivity of hydrides and the formation or destruction of the local magnetic moments in magnetic systems, etc. Here we break through the limitations of pressure on quantum sensors and develop the in-situ magnetic detection technique at megabar pressures with high sensitivity ($\sim$1 μT/$\sqrt{\text{Hz}}$) and sub-microscale spatial resolution. By directly imaging the magnetic field and the evolution of magnetic domains, we observe the macroscopic magnetic transition of $Fe_3O_4$ in the megabar pressure range from strong ferromagnetism (α-$Fe_3O_4$) to weak ferromagnetism (β-$Fe_3O_4$) and finally to non-magnetism (γ-$Fe_3O_4$). The scenarios for magnetic changes in $Fe_3O_4$ characterized here shed light on the direct magnetic microstructure observation in bulk materials at high pressure and contribute to understanding the mechanism of magnetic moment suppression related to spin crossover. The presented method can potentially investigate the spin-orbital coupling and magnetism-superconductivity competition in magnetic systems.

**Introduction**

Pressure has been proved to be powerful tools to tune the magnetic properties of materials as it can effectively increase the intermolecular interaction and redistribute the electrons [1-3]. For example, the spin crossover in 3d transition metal compounds depicted by the competition of the spin exchange and crystal field energy on compression[4], and the effect of covalency of the neighbored metal ions[5].

Additionally, measurements of magnetic properties at high pressures are extremely instrumental in high $T_c$ superconducting materials such as polyhydrides, where there is a high controversy about their existence and properties[6-8] due to a lack of sufficient evidence of the Meissner effect. Many efforts have been made to adapt magnetic detection methods to Diamond Anvil Cell (DAC) facilities. High energy photon/neutron scattering techniques such as x-ray magnetic circular dichroism (XMCD), Neutron magnetic scattering, and Mössbauer spectroscopy are widely used for magnetic measurements in DACs[9,10]. These techniques are able to probe atomic-scale magnetism and resolve local element-specific magnetism issues. However, these methods are not sensitive to magnetic phenomena like Meissner effect and technically restricted by sample and beam size at high pressure, and the quantitative analysis remains challenging on account of the resolution limited spectra[11]. The magnetization of the materials can be directly measured by integrating DAC into superconducting quantum interference devices (SQUID), but the effect of the DAC apparatus' background cannot be separated from the magnetization of the micrometer-scale samples, resulting in the lack of spatial resolution and the disadvantage to detect the magnetization of inhomogeneous sample synthesized at high pressure and high temperature inside the high-pressure chamber[12,13].

In recent years, NV centers in diamond have been proposed as quantum sensors for new type of high-pressure magnetic measurements, as they can be set in nanoscales to the sample inside the pressure chamber to detect the complex free-space magnetic field textures generated by high-pressure samples and able to provide significant potential advantages in terms of sensitivity and spatial resolution[14-16]. However, NV centers' magnetic sensing capabilities associated with their optical and spin properties could be dramatically degraded by the surrounding stress environment. Recent works show that in complex stress environment both the optical-spin readout contrast worsen and spin resonance linewidth becomes broader leading to a sharp drop in sensitivity as the pressure approaches megabar level[17,18]. Understanding the stress-induced effect and maintaining excellent performances of NV centers present substantial challenges for applying this technique to reveal the magnetically related physics above megabar pressures.

In this work, we explore the impact of stress on the optical and spin properties of NV centers. Then by modulating the uniaxial stress along nitrogen-vacancy axis, we improve the magnetic detective sensitivity to $\sim 1\ \mu T/\sqrt{Hz}$ at 130 GPa, which is promising for nano-scale single-domain grain detection[19,20]. Based on the sensitivity of the NV quantum sensors achieved, we investigated the magnetism change of the $Fe_3O_4$ – one of the oldest magnetic mineral on the earth[21] – at pressures well exceeding a megabar at room temperature.

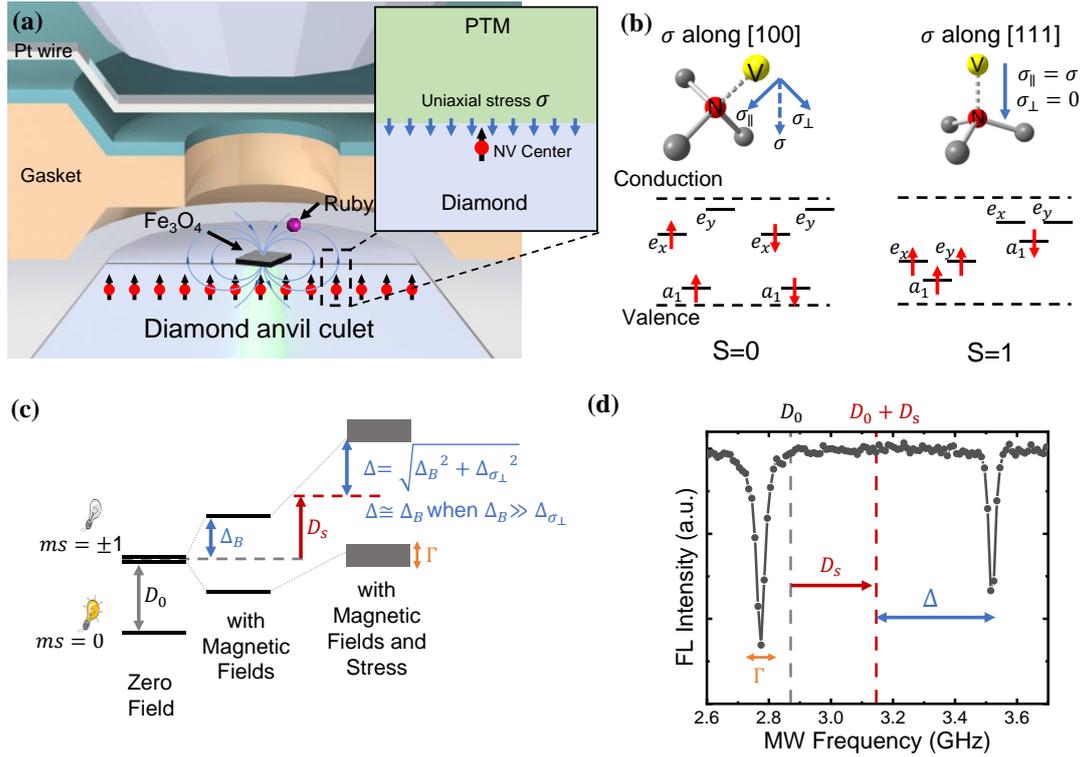

**Figure 1. Schematic geometry NV centers in the DAC. (a)** Schematic of the DAC geometry. The shallow layer of NV centers is embedded in the diamond anvil culet and a single crystal of $Fe_3O_4$ is placed on the surface of the same culet. The 532 nm laser is used to optically initialize and read out the spin state of NV centers. The Pt wire is used to transfer the microwave for controlling the spin state. Ruby fluorescence and diamond edge Raman are used to calibrate pressures in DAC chamber. The inset shows the uniaxial stress component σ perpendicular to the surface of culet. **(b)** The electronic ground state levels of the NV center under megabar uniaxial stress by first-principles calculations. Left: uniaxial stress applied to the [100] crystal direction. Right: uniaxial stress applied to the [111] crystal direction. **(c)** The spin ground state energy structure of the NV- and the evolution with the existence of both magnetic field and stress environment. The fluorescence intensity of the NV center is dependent on the spin states (ms=0, bright; ms = ±1, dark). And the spin states can be initialized to ms=0 by laser and coherently manipulated by microwave. $D_0$ = 2.87 GHz is the zero-field splitting. $\Delta_B$ is half of the Zeeman splitting energy with external magnetic field. $D_s$ is the energy shift of the zero-field splitting caused by the axial stress component. $\Delta_{\sigma_\perp}$ is half of the splitting in ms = ±1 spin sublevels produced by the shear stress in NV's frame. Γ is the additional broadening of the sublevels for the ensemble NV centers in inhomogeneous stress. **(d)** The corresponding optically detected magnetic resonance (ODMR) spectrum at 46.3 GPa presenting the spin states in (c). The dashed grey line marks the peak position ($D_0$) at ambient conditions with no magnetic field, and the dashed red line marks the central position ($D_0+D_s$) of two separated peaks at high pressure with a parallel magnetic field (~133Gs). The data extraction of $D_0$, $D_s$, Δ, and Γ allow to determine the magnitude of the stress and external magnetic fields.

## Results

**Quantum sensors at megabar pressures.**

Firstly, to understand the impact of the stress on the sensors, we investigated the electronic energy level of the NV center under uniaxial stress with different directions by first-principles calculations (Fig. 1b). The stress component $\sigma_\perp$ perpendicular to the NV axis breaks the $C_{3v}$ symmetry, leading to a ground spin state transition of NV from triplet state (S=1) to single state (S=0) and ultimately resulting in a loss of magnetic detection ability. However, the stress applied along the NV axis preserves the spin property of NV centers, enabling high-pressure magnetic measurements. At the anvil tip, the stress comprises both hydrostatic pressure (ρ) and uniaxial stress (σ) perpendicular to the surface[22], allowing for modulation of uniaxial stress direction on NV centers through control of anvil crystal orientation. For NV centers oriented along [111] direction in the (111)-cut anvil, the component $\sigma_\perp$ can be minimized. As a result, the contrast between ms=0 and ms = ±1 in the ODMR spectrum remains high as pressure increases to megabar range (Fig. 2a). Notably, at 130 GPa, we observed a contrast enhancement of ~30%, much higher than that at 1.5 GPa (Fig. 2f). This suggests that NV centers oriented in the other three crystallographically equivalent directions ([$1\bar{1}\bar{1}$], [$\bar{1}1\bar{1}$], [$\bar{1}\bar{1}1$]), which produce intense optical background signals at low pressure, no longer fluoresce at high pressure and thus increase the ODMR contrast (See supplementary information for details). In conclusion, the observed high contrast is approximately two orders of magnitude higher than previous results at similar pressures [23,24].

Secondly, we investigate the effect of inhomogeneous stress on the broadening of the ODMR spectrum (Γ). Together with the ODMR spectrum contrast (C), these two parameters are crucial in determining the sensitivity of magnetic measurements[25,26]:

$$\eta_B = \frac{4}{3\sqrt{3}} \frac{h}{g_e \mu_B} \frac{\Gamma}{C\sqrt{R}}$$

with the Planck constant $h$, g-factor $g_e$, Bohr magneton $\mu_B$ and photon-detection rate R.

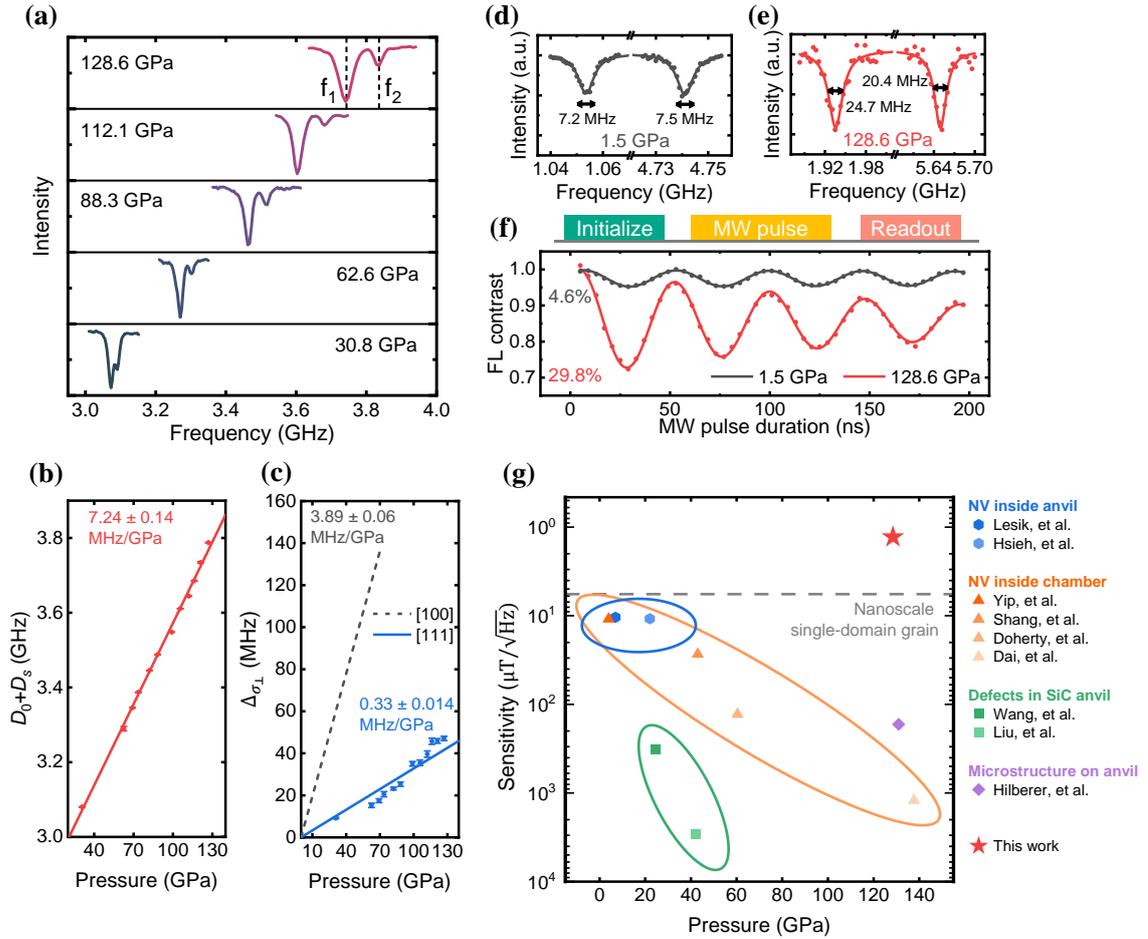

**Figure 2. NV centers in crystal orientation designed diamond anvil (a)** Zero magnetic field ODMR spectrum of NV centers in pressure, $f_1$ and $f_2$ are the resonant frequencies of the left and right peaks in the spectral line respectively, ODMR center frequency $D_0 + D_s = (f_2 + f_1)/2$, and ODMR splitting induced by shear stress $2\Delta_{\sigma_\perp} = f_2 - f_1$. **(b)** Pressure dependence of $D_0 + D_s$. The data are represented by red filled circles with error bars, and the red solid line represents the linear fitting of the data, showing a linear shift of 7.24 MHz/GPa **(c)** Pressure dependence of $\Delta_{\sigma_\perp}$. The blue filled circles with error bars represent the data in (111)-cut anvil, and the blue solid line represents the linear fitting of the data. The dashed grey line represents the result in <100>-cut anvil by Hilberer et al. showing a linear shift of 0.33 MHz/GPa in (111)-cut anvil and 3.89 MHz/GPa in <100>-cut anvil[24]. **(d)** and **(e)** The ODMR spectrum linewidth (Γ) tested in ~660 Gs magnetic field at 1.5 GPa and 128.6 GPa. The spectrum broadening is about 7 MHz and 20 MHz at the pressure of 1.5 GPa and 128.6 GPa, respectively. **(f)** Rabi oscillations of NV electron spins. The ODMR contrast is 4.5% and 29.8% at the chamber pressures of 1.5 GPa and 128.6 GPa, shown by gray and red filled circles respectively. The experimental data are fitted by damped sine wave function and plotted by the gray and red solid lines. **(g)** Comparison of high-pressure magnetometry techniques based on color centers in diamond anvil cell. Our work is marked by red star. The strategy of using NV inside the (100)-cut diamond anvil is marked in blue[14,15]. The strategy of using NV inside the pressure chamber is marked in orange[16,17,23,27]. The strategy of fabricating microstructure on anvil is marked

in violet[24]. The strategy of using the defects in SiC anvil is marked in green[28,29]. The dashed grey line indicates the magnitude of the remnant magnetic field (~5.8 µT) generated by a nanoscale single-domain grain (magnetic moment $10^{-17}$ A·m$^2$, distance 700 nm).

As shown in Fig. 1c, d and Fig. 2b, in the presence of axial stress, the center frequency of ms = ±1 spin sublevels shifts from $D_0$ (2.87 GHz) to $D_0 + D_s$[30,31]. In addition, inhomogeneous axial stress along the NV axis also broadens the ODMR spectrum and reduces the magnetic measurement sensitivity. However, since the stress gradient is mainly perpendicular to the anvil surface and the distribution of the NV centers in z direction is highly concentrated (~4nm thickness through low-energy (9keV) ion implantation) we can largely suppress the linewidth broadening. As shown in Fig. 2d and e, the linewidth of the ODMR spectrum is narrowed to approximately 20 MHz, a five-fold reduction compared to previous work[23,24]. With an external magnetic field projected along NV axis ($B_z$) and shear stress ($\sigma_\perp$), the energy splitting between ms = ±1 sublevels can be expressed as $2\Delta = 2\sqrt{\Delta_B^2 + \Delta_{\sigma_\perp}^2}$, determined by the Zeeman splitting ($2\Delta_B = 2\gamma_e B_z$) and the shear stress splitting ($2\Delta_{\sigma_\perp}$). In this work, we minimize the shear stress using a (111)-cut anvil and obtain $\Delta \cong \Delta_B$ since $\Delta_B \gg \Delta_{\sigma_\perp}$ (Fig. 2c).

As shown in Fig. 2g, the improvement of ODMR contrast and the suppression of spectrum broadening result in a magnetic detection sensitivity of ~1 µT/√Hz at megabar pressure. and the sensor has a sub-microscale spatial resolution due to the combination with a confocal optical detection system (objective NA = 0.4). With such detection sensitivity and resolution, our quantum sensors can investigate the remnant magnetic field generated by nanoscale single-domain grains[19,20] at megabar pressures.

**Magnetic properties investigation at megabar pressures.**

Using the quantum sensors, we studied the evolution of the magnetic properties of $Fe_3O_4$ at pressure from ambient to megabar levels. As the mechanism of the magnetic transition remains unclear, the suppression of the exchange interaction and the spin-orbital coupling is especially complicated in the case of mixed Fe ion valence. Critically, magnetic data on a worldwide mineral collection reveal that pure magnetite which is common in the uppermost mantle and haematite could carry a magnetic remanence down to ~600 km[21]. The studies on its high-pressure magnetic behaviors are important not only for the fundamental physical and chemical science but also for the understanding of the geomagnetism. Very recently, the single crystal experimental study on Fe$_3$O$_4$ clarified the structure and established the phase diagram at extreme conditions[32]. This covers three high-pressure phases of Fe$_3$O$_4$: α-Fe$_3$O$_4$ ($Fd\bar{3}m$), β-Fe$_3$O$_4$ (Bbmm) and γ-Fe$_3$O$_4$ (Pbcm) and the structure changes were related to spin crossover, which have been extensively investigated by neutron scattering, Mossbauer spectroscopy, Fe K-edge X-ray absorption, and magnetic circular dichroism measurements techniques at high pressure[9,10,33-35]. These experimental results are somewhat contradictory possibly due to the inhomogeneity of the samples such as oxygen vacancies, a variety of stresses induced by different pressure medium, and low resolution of the spectrum at high pressures. Moreover, these experiments studies, which investigate the spin cross over at different Fe sites in the atomic scale, lack experimental observations of the macroscopic magnetism of the samples. A complete understanding of macroscopic effects is almost impossible without knowledge of the underlying domain structure[36,37].

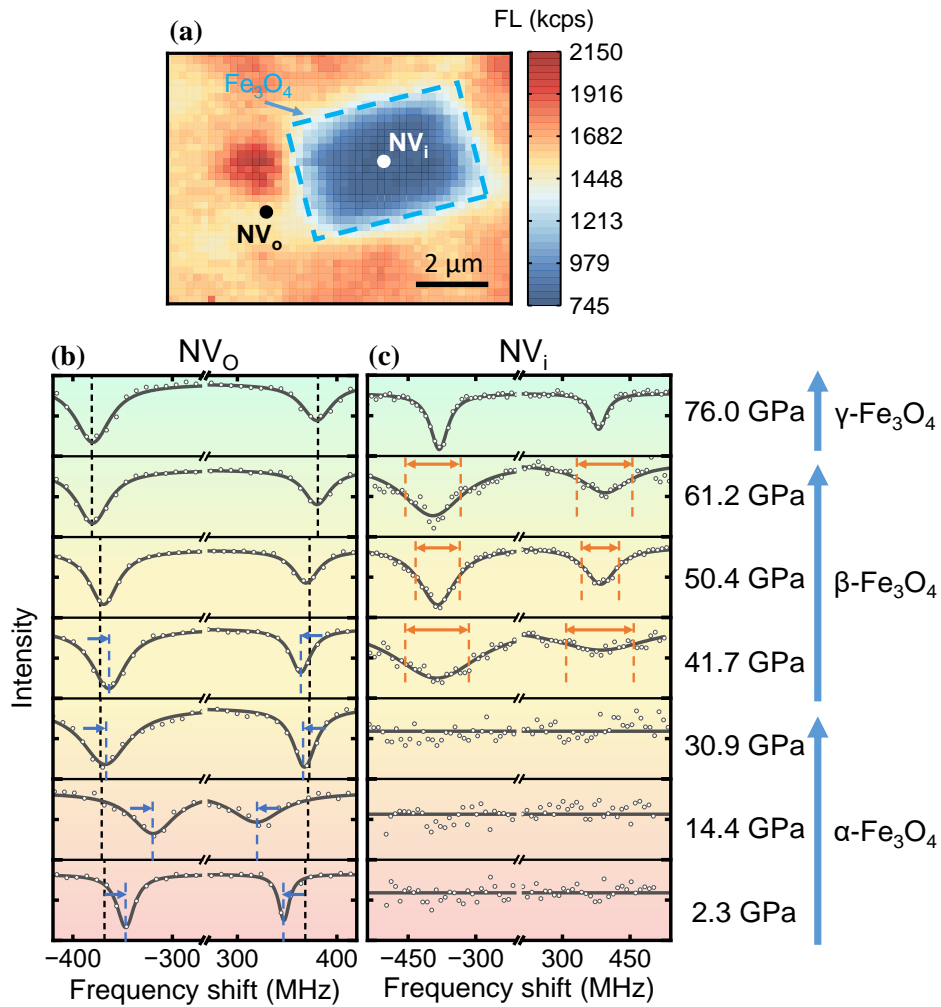

**Figure 3. Data for magnetic detection of magnetite (a)** Fluorescence (FL) imaging of NV centers in the anvil tip. Due to the fluorescence resonance energy transfer effect, FL below $Fe_3O_4$ sample is reduced. The dashed blue square represents the area of $Fe_3O_4$ sample for eye guide. The filled black and white circle marks the NV centers below the side ($NV_o$) and beneath ($NV_i$) the $Fe_3O_4$ sample, respectively. **(b)** and **(c)** The ODMR spectra of $NV_o$ and $NV_i$ change with elevated pressures, respectively. The horizontal axis represents the frequency shift relevant to the center frequency of ms = ±1 spin sublevels. The dashed black lines mark the resonant frequency of the NV centers far away from the sample, which is used to characterize the magnitude of the external magnetic field along NV axis (~130 Gs). The dashed blue lines and arrows mark the resonant frequency shift induced by the magnetic field from the sample. The dashed orange lines and arrows mark the spectrum broadening induced by magnetic field fluctuations from the sample.

Here, we carry out the study on the macroscopic magnetism of $Fe_3O_4$ under megabar pressure. The pressure dependence of magnetic properties is characterized by the ODMR spectra of $NV_O$ and $NV_i$, as shown in Fig. 3b and c, respectively. The positions of $NV_O$ and $NV_i$ are referred to the $Fe_3O_4$ sample as illustrated in Fig. 3a.

For $NV_O$, the static magnetic field induced by magnetite results in additional splitting relative to an external constant bias magnetic field (Fig. 3b). The pressure dependence of the static magnetic field is shown in Fig. 4a, overall revealing a trend of first rising and then falling, and afterwards approaching to 0. From 0-20 GPa, we observe a change in magnetocrystalline anisotropy due to stress. Considering that the external magnetic field is along <110> of $Fe_3O_4$ and the initial easy axis is <111> at ambient pressure, this change in magnetocrystalline anisotropy results in a magnetization enhancement from 0 GPa to 15 GPa. In 20-30 GPa pressure range, magnetic discontinuities and reversals due to drastic changes in magnetic anisotropy are observed, which is close to the first structural phase transition point. After entering the β-$Fe_3O_4$ phase (30-65 GPa), the magnetic field induced by magnetite gradually degrades, and finally leads to a disappearance of the extra splitting at around 70 GPa.

For $NV_i$, the magnetic fluctuations on magnetite surface results extra linewidth broadening of ODMR spectra (Fig. 3c). The pressure dependence of the linewidth broadening is shown in Fig. 4b. However, in the α-$Fe_3O_4$ phase (0-30 GPa), due to the strong transverse magnetic field ms = ±1 and ms =0 state mixes with each other leading to the disappearance of the ODMR contrast (Fig. 3c)[38]. As this mixing effect diminishes with the weakening of magnetic field in the β-$Fe_3O_4$ phase, the expected ODMR is observed but with a strong linewidth broadening due to magnetic fluctuations near the sample. The emergence of magnetic fluctuations implies the existence of intense frequent movement of the domain walls[39], driven thermally or magnetically[40], or probably stimulated by microwaves or laser light[41]. As the pressure increases above 70 GPa, the ODMR broadening caused by magnetic fluctuations disappears (Fig. 4b), implying the disappearance of magnetic domains into the γ-$Fe_3O_4$ phase, with a high probability associated with the structural phase transition (see supplementary information for phase transition XRD measurement).

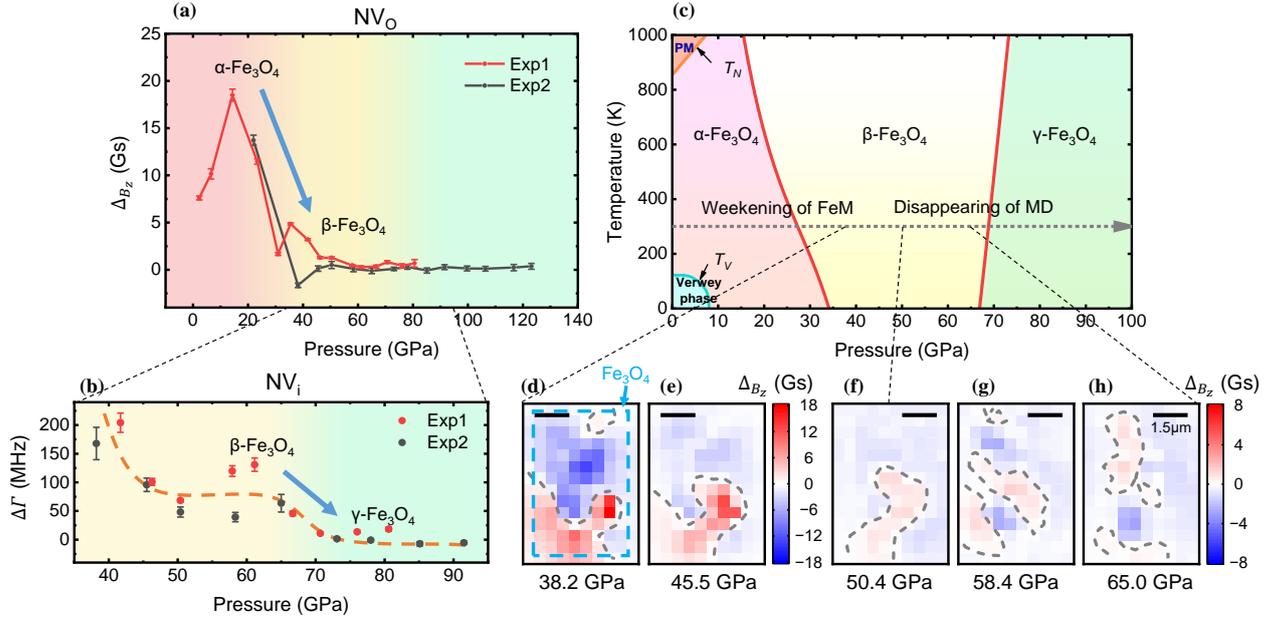

**Figure 4. Demagnetization of magnetite to megabar. (a) and (b)** Pressure dependence of the extra magnetic field $\Delta_{B_z}$ at $NV_o$ and the extra ODMR linewidth $\Delta_\Gamma$ of $NV_i$, respectively. $\Delta_{B_z}$ is magnetic field at $NV_o$ minus the external magnetic field. $\Delta_\Gamma$ is ODMR spectrum linewidth of $NV_o$ minus the inherent linewidth. The external magnetic field and the inherent linewidth are obtained from the NV centers far away from the sample. The red and black dots with error bars show the experimental results from two DACs with an external magnetic field of ~130 Gs and ~570 Gs along NV axis respectively. **(c)** The phase diagram of $Fe_3O_4$[10]. **(d) to (h)** Magnetic field imaging of the surface of $Fe_3O_4$ at pressures of 38.2, 45.5, 50.4, 58.4 and 65.0 GPa, respectively. The dashed blue line in (d) marks the $Fe_3O_4$ sample and dashed gray lines mark the magnetic domain wall.

The magnitude of the static magnetic field (Fig. 4a) and the magnetic fluctuations (Fig. 4b) reveal that, with increasing pressure, the magnetite undergoes a weakening of ferromagnetism (FM) (α-$Fe_3O_4$ to β-$Fe_3O_4$) and an eventual loss of ferromagnetism (β-$Fe_3O_4$ to γ-$Fe_3O_4$). Imaging the magnetic domains (MD) further enriches the details of this process. In the pressure range of 38 to 50 GPa (Fig. 4d to f), the shape of the MD does not change significantly, but the magnetization intensity decreases, indicating that the process remains in the ferromagnetic phase and is accompanied by a weakening of the ferromagnetism. In the pressure range of 50 GPa to 70 GPa (Fig. 4f to h), the MD experience a drastic change in shape, implying that the process of MD disappearance begins to occur before the γ-phase transition (70 GPa).

**Discussion**

Our work extends magnetic measurement technique based on NV centers to the megabar pressure range. Utilizing the improved performance of the quantum sensors, we substantially detect the magnetic properties of single crystal $Fe_3O_4$ under megabar pressure. Our high spatial resolution magnetic measurement technique can overcome sample inhomogeneities and reveal micro-nanometer scale local magnetism evolution. As demonstrated in our experiment, we observed the evolution of magnetic domains in samples through magnetic imaging.

As illustrated in Fig. 4a, the abrupt drop of $\Delta_{B_z}$ above 20 GPa implicates the weakening of magnetism, which means that the spin transition is likely occur in advance of structural transition to β-$Fe_3O_4$. The $\Delta_{B_z}$ at around 40 GPa from two runs of experiments with positive and negative values indicates change in magnetic anisotropy, owing to the magnetic transition related to α-$Fe_3O_4$ to β-$Fe_3O_4$. It is worthy to note that the ferromagnetic phase sustained above 40 GPa and below ~420 K was observed by Mossbauer spectroscopy[10]. Meanwhile, there is DFT calculations showing that β-$Fe_3O_4$ (Bbmm) possesses smaller net magnetic moment of $1.7\,\mu_B/f.u.$ compared with $4.0\,\mu_B/f.u.$ of α-$Fe_3O_4$ phase[4]. Our measurements (Fig. 4d to h) do show that very low magnetization in β-$Fe_3O_4$ but still has magnetic domains, the intensity and shape of domain wall changes above 50.4 GPa (Fig4. f), which might correlate to the onset of magnetic phase transition. This is corresponding to the spin crossover, likely in partial cites of Fe cations, as reported in octahedral sites of $Fe^{3+}$[34]. Furthermore, the disappearance of the magnetic domains in γ-$Fe_3O_4$ implies a possible entry into non-magnetic state. The possible reason is that only $Fe^{3+}$ ions occupying square antiprism coordinated sites still remain in the HS state (above 84 GPa and down to 2.4 K), which interrupt the magnetic interaction between $Fe^{2+}$ and $Fe^{3+}$ components, suggesting a nonmagnetic state rather than paramagnetic state[34]. However, the Fe K-edge X-ray absorption and magnetic circular dichroism measurements suggest the loss of the net ordered magnetic moments in β-$Fe_3O_4$ above 33 GPa, indicating the low spin in three different sites of Fe ions and, thus, a nonmagnetic state[9]. Therefore, in the future work, we expect to combine the variable temperature and magnetic field to obtain the magnetic P-T phase diagram and saturation magnetization of magnetic materials.

This technique is also applicable to in-situ direct detection of high-temperature superconducting materials under megabar pressure, including various superhydrides compounds[7,42]. In addition to the DC magnetic field measurements, the optical and spin properties of our NV centers also lift the limitations of nuclear magnetic resonance (NMR) and electron spin resonance (ESR) measurements under megabar pressures. Thus, this work is promising to enabling magnetic resonance imaging (MRI) in pressure chambers thus providing an opportunity to make a great impact generally in investigating high-temperature superconductivity at high pressures [6,7,43-45].

**Methods**

In our experiments, we use a BeCu symmetric diamond anvil cell (Fig. 1a) compressing rhenium gasket to provide a high-pressure environment for $Fe_3O_4$ samples. A platinum wire is compressed between the gasket and anvil pavilion facets serves as the microwave radiation guide. Nano cBN mixed with epoxy powder isolates the Pt foil and gasket. The sample chamber confined by the gasket and diamond culets is filled with pressure-transmitting medium (PTM) KCl to provide the quasi-hydrostatic environment. The magnetite sample is placed on the diamond surface and compressed together with PTM. Ruby fluorescence (below 20GPa) and Raman spectra of the diamond (above 20 GPa) are used as pressure calibrants[46].

Diamond anvil we used in experiments are cut and polished from HPHT type-IIas (Non-fluorescent) single crystal diamonds. Both 100 μm and 150μm diameter the culets have been used. The NV centers layer used for magnetic sensing is positioned about ~9 nm deep from the surface of the anvil. The NV centers layer is create by $^{14}N^+$ ion implantation (Energy: 6 keV Dose: $1*10^{13}/cm^2$ ) and vacuum annealing treatment (1000℃ ~2h). The size of the magnetite sample is about 4 μm *5 μm*1 μm, which is cut from a bulk 99.99% purity single crystal $Fe_3O_4$ (HeFei Crystal&Surface Technical Material Co., Ltd).

A home-built optical detected magnetic resonance system is used to address NV ensembles inside the DAC and collect fluorescence signal under extreme pressure. (Fig. S1) We use a 532nm laser (CNILaser MGL-III-532-150mW), controlled by two serial acousto-optic modulators (AOM Gooch&Housego AOMO 3200-121), to initialize NV ensembles. The laser beam is focused on the anvil tip through the light port of the DAC using a long working distance objective lens (Mitutoyo MY20X-824, NA0.4), which is mounted on a piezo scanning stage (Coremorrow P12A.XYZ100S). Through the same objective lens, the NV fluorescence is collected and a dichroic mirror is used to separate it from excitation laser beam. A flip mirror is used to switch the light path between a spectrometer (Horiba iHR550) for pressure measurement and the confocal system, which comprises a 50um pinhole and a 633nm long-pass filter (Semrock BSP01-633R-25). A single photon counting module (Excelitas SPCM-780-24-BR1) is used to collect the fluorescence signal. We use a counter card (National Instruments PCIe-6612) for photon signal counting. A microwave source (Rigol DSG3060) combining a 50W amplifier (Mini-Circuits ZHL-50W-63+) serves to generate MW signal for NV spin manipulation. A Pulse Generator (CIQTEK ASG8100) is used to output pulse sequence and provide timing control for the whole system.

**Acknowledgments:** The fabrication of diamond sensors was partially performed at the USTC Center for Micro and Nanoscale Research and Fabrication, and the authors particularly thank Dr. Xiaolei Wen, Xiwen Wang, Cunliang Xin, and Hongfang Zuo for their assistance in $Fe_3O_4$ and sensors fabrication process and Dr. Elena Bykova for her beamtime with X-ray diffraction measurements of $Fe_3O_4$ at high pressure. This work is supported by the National Natural Science Foundation of China (Grants No. 92265204, 12104447, 12204485), the CAS (GJJSTD20200001), the National Key R&D Program of China (Grants No.2021YFB3202800), the Innovation Program for Quantum Science and Technology (Grant No. 2021ZD0302200), the Anhui Initiative in Quantum Information Technologies (Grant No. AHY050000), the Fundamental Research Funds for the Central Universities.

**Author contributions:** Jiangfeng Du, Ya Wang and Yu Wang supervised the project, Ya Wang, Yu Wang, and Mengqi Wang proposed the idea of the experiment. Mengqi Wang, and Zhixian Liu built the experimental set-up and performed the measurements. Yu Wang and Ganyu Xu prepared the diamond anvil cell integrated quantum sensor. Mengqi Wang prepared the NV center and the magnetite sample with the help of Pei Yu, Haoyu Sun, and Xiangyu Ye. Bo Yang performed a first-principles calculation. performed the data analysis with contributions from all co-authors. Mengqi Wang, Yu Wang, Zhixian Liu, and Ya Wang performed the data analysis and wrote the manuscript with contributions from all co-authors. All authors contributed to the discussion of the results. All authors discussed the results and commented on the manuscript.

**Competing interests:** The authors declare that they have no competing interests.

**Corresponding author:**

Jiangfeng Du(djf@ustc.edu.cn), Ya Wang(ywustc@ustc.edu.cn) and Yu Wang (wangyu@issp.ac.cn)